\documentclass[prb,onecolumn,amsmath,amssymb]{revtex4}

\usepackage{graphicx}
\usepackage{dcolumn}
\usepackage{bm}
\usepackage{epstopdf}

\def\be{\begin{equation}}
\def\ee{\end{equation}}
\def\bd{\begin{displaymath}}
\def\ed{\end{displaymath}}
\def\-{\phantom{-}}

\begin{document}

\title{Proximity-induced magnetism in transition-metal substituted graphene}


\author{Charles B. Crook$^1$}
\author{Costel Constantin$^1$}
\author{Towfiq Ahmed$^2$}
\author{Jian-Xin Zhu$^{2,3}$}
\author{Alexander V. Balatsky$^{4,5}$}
\author{Jason T. Haraldsen$^1$}
\affiliation{$^1$Department of Physics and Astronomy, James Madison University, Harrisonburg, VA 22802}
\affiliation{$^2$Theoretical Division, Los Alamos National Laboratory, Los Alamos, NM 87545, USA}
\affiliation{$^3$Center for Integrated Nanotechnologies, Los Alamos National Laboratory, Los Alamos, NM 87545, USA}
\affiliation{$^4$Institute for Materials Science, Los Alamos National Laboratory, Los Alamos, NM 87545, USA}
\affiliation{$^5$NORDITA, Roslagstullsbacken 23, 106 91 Stockholm, Sweden}

\date{\today}

\begin{abstract}

We investigate the interactions between two identical magnetic impurities substituted into a graphene superlattice. Using a first-principles approach, we calculate the electronic and magnetic properties for transition-metal substituted graphene systems with varying spatial separation. These calculations are compared for three different magnetic impurities, manganese, chromium, and vanadium. We determine the electronic band structure, density of states, and Millikan populations (magnetic moment) for each atom, as well as calculate the exchange parameter between the two magnetic atoms as a function of spatial separation. We find that the presence of magnetic impurities establishes a distinct magnetic moment in the graphene lattice, where the interactions are highly dependent on the spatial and magnetic characteristic between the magnetic atoms and the carbon atoms, which leads to either ferromagnetic or antiferromagnetic behavior. Furthermore, through an analysis of the calculated exchange energies and partial density of states, it is determined that interactions between the magnetic atoms can be classified as an RKKY interaction. 

~

\noindent Corresponding Author: Dr. Jason T. Haraldsen (haraldjt@jmu.edu)


\end{abstract}
\maketitle

\section*{Introduction}

For the last century, the vast increase in technological capabilities has been largely governed by the ability of devices to manipulate and control charge in various materials. However, charge is not the only property of electrons that can be exploited for technological use. Within the last decade, there has been a surge in the number of devices and materials that work to control, not only charge, but also the electron spin degrees of freedom. This evolution has brought about the area of spintronics\cite{wolf:01,bade:10,zuti:04,roch:05,boga:08,khaj:11}, where devices made from spintronic materials are being investigated for applications ranging from quantum computing to enhanced nano-sized memory storage\cite{khaj:11,enge:01,awsc:02}, because they typically use less power and provide a dramatic increase in computer storage capacity and capability. 

The use for spintronic devices varies dramatically due to the multitude of interesting phenomena they exhibit. This includes spin relaxation and transport that may be used for quantum computation\cite{khaj:11,enge:01,awsc:02}. However, there are distinct challenges for spintronics, and a great need to completely understand the device applications for these materials \cite{awsc:07}. Therefore, efforts in areas of experimental and theoretical physics, engineering, and chemistry are working to identify materials for spintronic applications\cite{xu:06,mats:03}, as well as working toward understanding the fundamental interactions that govern these material properties\cite{sato:02}. One large area of interest for these materials is that of two-dimensional (2D) materials\cite{wolf:01,bade:10,xiao:12}, where flat (or nearly-flat) organizations of atoms display extraordinary surface properties and may be useful for smaller and more accessible electronics\cite{saha:10}. 

\begin{figure}
\includegraphics[width=0.35 \linewidth]{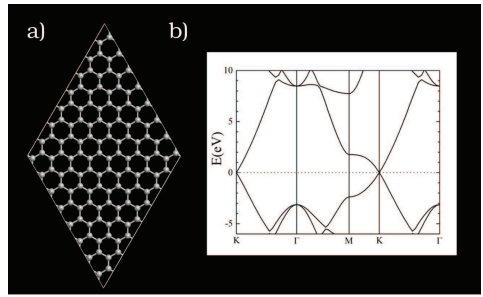}
\caption{(Color Online) a) The 2D hexagonal structure of graphene and b) the calculated electronic band structure.}
\label{gen-sub}
\end{figure}

The most popular example of a 2D material is that of graphene\cite{boehm:62,geim:07}, which exhibits a large electronic mobility and thermal conductivity\cite{neto:09,chen:08,pop:12,bolo:08,moro:08} that provides the ability to conduct electrons. Recently, graphene has gained considerable attention in the scientific community due to the technological applications in electronic circuits and possible memory devices\cite{saha:10,chen:08}, which can allow for better transistors for faster processing and flexible circuits.

Graphene consists of an individual layer of carbon atoms arranged in a standard honeycomb lattice (Fig. \ref{gen-sub}(a))\cite{wehl:14,novo:05,zhu:10}. This configuration allows for graphene to have a large tensile strength and ability to conduct electrons through various $p$-orbital ($\pi$) bonding\cite{lee:08}. Therefore, the tensile strength of graphene, relative to its weight and thickness, has been shown to be stronger than steel\cite{sava:12}, which makes it a durable material for possible use in nanotechnological devices\cite{ahme:14}. Furthermore, graphene has a specific crossover point in its electronic band structure, called the Dirac point, that occurs at the $K$-point of the Brillouin zone (Fig. \ref{gen-sub}(b)), and provides the system a higher electron mobility.  Because graphene is essentially a manifestation of surface states, it is considered to be a strong 2D topological Dirac material. 

Since graphene is being investigated as a base material for electronics, the current challenge is to establish it for use as a spintronic material. However, graphene is not magnetic by itself\cite{yazy:08,rao:12}. Therefore, the ability to use graphene as a spintronic material is not very clear\cite{pesi:12}. There have been a number of theoretical studies that have examined the viable of substituting magnetic atoms into graphene\cite{sant:10,sher:11,hu:11}. Recently, an experimental study showed that graphene deposited on an yttrium iron garnet thin film can induce a ferromagnetism in the graphene layer\cite{wang:15}. However, the mechanism that governs these interactions needs to be clarified.

Another mechanism for possible magnetic induction in graphene is through direct magnetic substitution, where a carbon atom is removed from the graphene lattice and replaced with a magnetic atom. This can provide the necessary exchange energy and help clarify the underlying mechanisms. It is possible to provide direct substitution of magnetic atoms using scanning tunneling microscopy methods. The process of removing individual carbon atoms from the graphene lattice was recently demonstrated by Gomes et al.\cite{gome:12}. Furthermore, direct substitution of nitrogen and cobalt has been observed\cite{wang:12,lv:12}. Therefore, it is viable that magnetic atoms could be placed in the graphene lattice. This will be discussed further below.

\begin{figure}
\includegraphics[width=0.35 \linewidth]{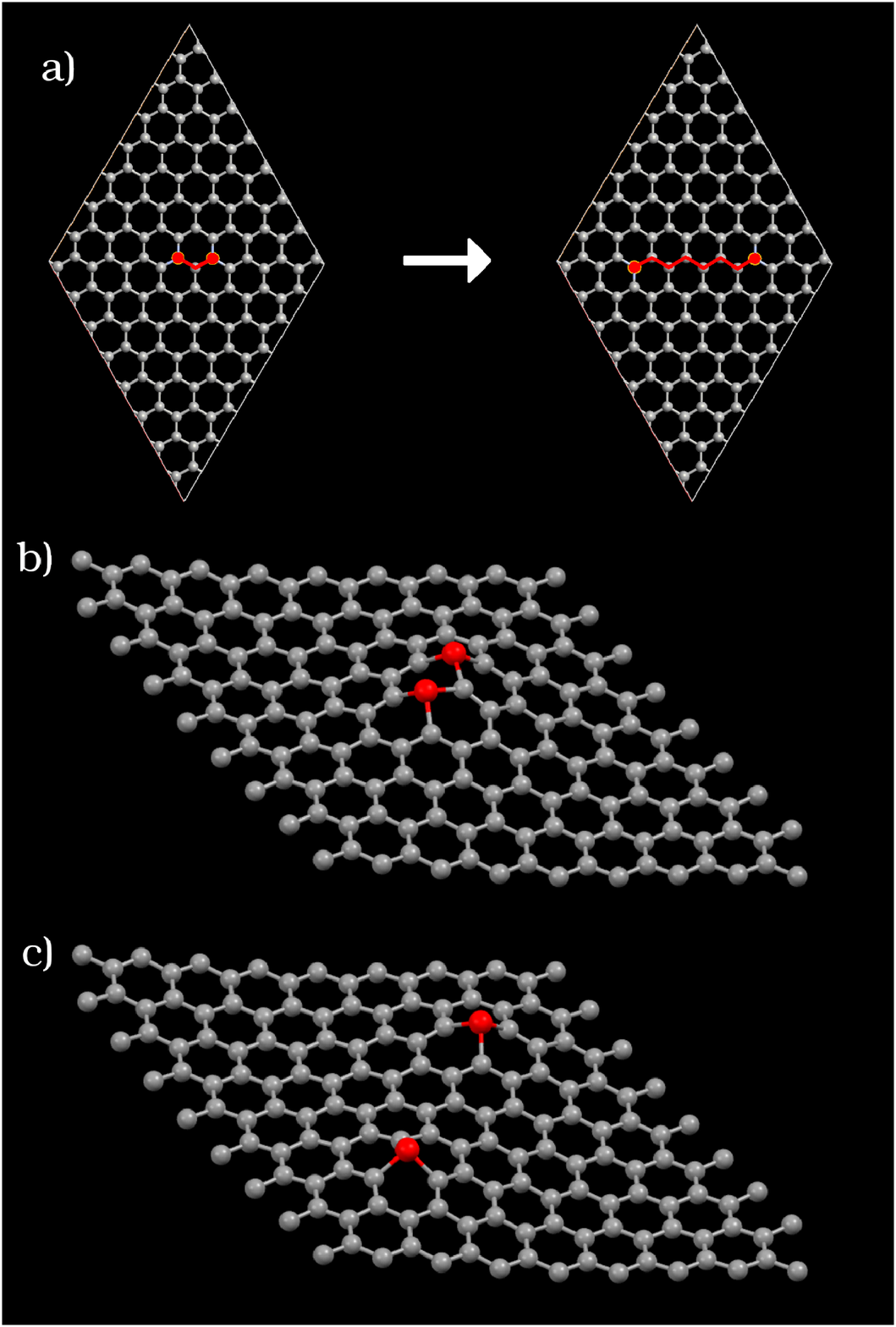}
\caption{(Color Online) a) The general substitution positions. 3-D view of the distorted graphene structure with two magnetic impurities separated by 1 carbon (b) and 6 carbons (c).}
\label{gen-dis}
\end{figure}

In this article, we present a comparative computational study examining the changes in the electronic and magnetic properties of transition-metal substituted graphene. Starting from a 128-atom supercell of graphene, two carbon atoms are replaced with either vanadium, chromium, or manganese impurities. The magnetic impurities are considered for various spatial separations of N carbon atoms (N = 1 - 6) in between the magnetic atoms along the zig-zag chain direction in graphene. Using density functional calculations with a full geometry optimization and spin polarization, we performed calculations to determine the total energy ground state, electronic band structure, density of states (DOS), and magnetic moment for each spin configuration. Overall, it is determined that the presence of magnetic atoms induces a magnetic state in the graphene, where the partial density of states shows a distinct increase in electron mobility at the Fermi level producing a metallic state. Additionally, we map the magnetic moment for each atom in superlattice to examine the magnetic exchange (or super-exchange) interaction characteristics throughout the graphene layer. Overall, this study provides clarity on how the magnetic atoms are communicating through graphene layer as well as helps in the understanding of the super-exchange interactions between the impurity atoms.

Furthermore, through an analysis of the calculated exchange energy between the magnetic atoms, we find that the interaction alternates between a ferromagnetic and antiferromagnetic state as the magnetic atoms are separated spatially. This, in combination with the metallic state in the carbon, illustrates and suggests the presence of an RKKY exchange interaction. An RKKY (Ruderman-Kittel-Kasuya-Yosida) interaction is an indirect exchange coupling that is governed through interaction of localized spin moment and conduction electrons\cite{rude:54,kasu:56,yosi:57}. This produces two main criteria: 1) an alternating or oscillatory exchange strength dependent on distance, and 2) a distinct conduction of electrons typically in a metallic state. Therefore, this study provides avenues for the electronic and magnetic manipulation of graphene that can be utilized in the realization of spintronic and quantum computing devices.

\begin{figure}
\includegraphics[width=0.4 \linewidth]{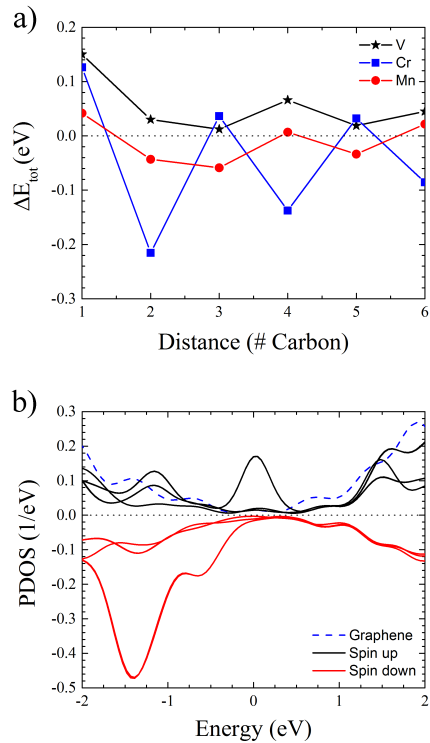}
\caption{(Color Online) a) Calculated exchange energy (change in total energy) for vanadium (black stars), chromium (blue squares), and manganese (red circles) substituted graphene. b) The partial density of states for the connector carbons (positive - spin up channel and negative - spin down channel), as well as that of pure graphene (dashed blue). Since this is for the five carbon case, there are three spin up and three spin down curves. This is because there are two pairs of carbons atoms that are spatially symmetry and therefore equivalent.}
\label{exch}
\end{figure}

\section*{Results}

We conducted simulations on magnetically-substituted graphene lattice to see how the presence of magnetic atoms would affect the electronic and magnetic structure in the system. Therefore, we replaced two carbon atoms with vanadium, chromium, or manganese in a symmetric dimer configuration. Through DFT simulation of the formation energy, it was determined that graphene with magnetic atom substitution is energy favorable over a pure graphene state. To examine the formation energy, we determined the total energy for a pure graphene sheet with a unbound magnetic atom (total energy $\sim$ -288 eV) and compared that to a graphene sheet with a bound magnetic atom and a free carbon (total energy $\sim$ -296 eV). Here, it is found that the magnetic substitution is energy favored by about 8 eV. This is most likely due to the number of electrons that can be shared by the magnetic atoms over the carbon.

To examine the electronic and magnetic interactions, we started with two atoms separated by a single carbon atom, and then separating them by one carbon along the zig-zag chain of graphene until there were six carbon atoms between the magnetic atoms (Fig. \ref{gen-dis}(a)). This allows us to examine the spatial dependence of the magnetic interaction between the two impurities. In order to magnetically map the supercell, we calculate the Mulliken population over the entire lattice. Furthermore, we determine the electronic band structure and the density of states (DOS) for comparison to that of pure graphene (Fig. \ref{gen-sub}(a) and (b)).

\subsection*{Impurity distortion}

The presence of the magnetic atoms produce a distinct distortion in the graphene lattice. This distortion is shown in Fig. \ref{gen-dis}(b) and (c), where the three-dimensional view of the cell with 1 carbon and 6 carbon respectively. Through a geometry optimization, it is determined that the presence of magnetic atom impurities produces a dramatic distortion in the graphene lattice, which is consistent with previous first-principles calculations\cite{sant:10} and experimental measurements \cite{neto:09,sing:09}. From this study, it can be determined that the bond lengths around the magnetic atoms are increased from the standard C-C length of 1.42 \AA~ to 1.87 \AA. This produces a distortion in the magnetic atom position and reduces the C-M-C bond angle from 120$^{\circ}$ to 92.9$^{\circ}$. This is due to the destabilizing presence of the larger magnetic atoms. Further information on the distortion can be found in the supplementary material.

\subsection*{Exchange energy and density of states}

Figure \ref{exch}(a) shows the exchange energy between the magnetic atom in each of the six configurations. The exchange energy is determined by examining the difference between the DFT calculated total energies for the FM and AFM states. According with the model, if the exchange energy is positive, then the interaction is FM. If the exchange energy is negative, then it is AFM. As shown in Fig. \ref{exch}(a), all magnetic substitutions show a distinct RKKY variation.  Although, only chromium and manganese fluctuate between FM to AFM, the vanadium interaction is constantly ferromagnetic and oscillates in the exchange strength. It should be noted that the magnetic ordering temperature is related to the strength of exchange coupling, where in this case, the ordering temperature of a spin dimer (FM or AFM) is dependent on the distance between the two magnetic atoms. 

Focusing mainly on chromium, Fig. \ref{exch}(b) shows the local density of states (LDOS) for the connector carbons in the chromium case (solid) as well as a pure graphene sheet (dashed). The DOS for all configurations is given in the supplementary material. Note that the density of states of pure graphene is zero at the Fermi level, while transition-metal substituted graphene has a distinct increase in the electron density of states at the Fermi level. This is a fundamental feature of Dirac materials where the presence of an impurity resonance at the Fermi level is produced as a gap forms in the Dirac cone\cite{wehl:14}.This provides a metallic DOS at the Fermi level that enables the conduction electrons to couple to the magnetic moment producing a super-exchange between the chromium atoms. This increase in the PDOS shows that the magnetic moments on the impurity atoms are influencing the electronic character of carbons between the magnetic atoms. Therefore, it is evident that the magnetic super-exchange is likely mediated through an RKKY coupling.

\subsection*{Magnetic Induction}

\begin{figure}
\includegraphics[width=0.4 \linewidth]{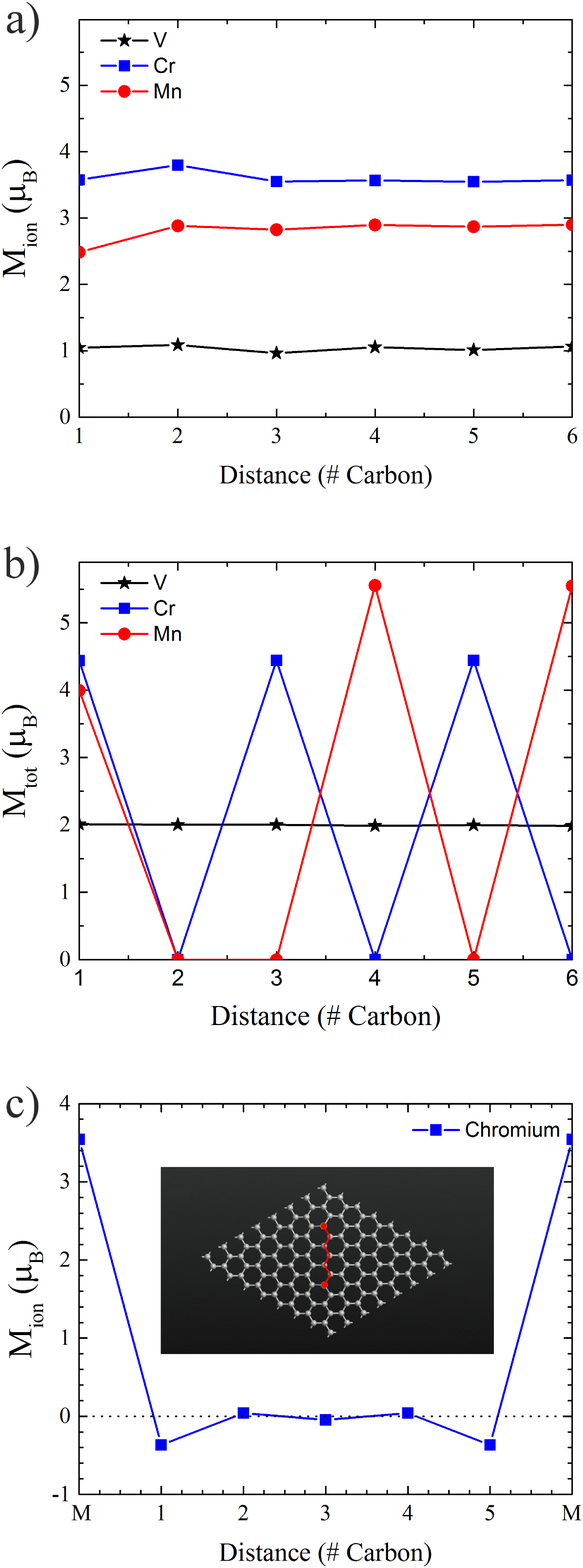}
\caption{(Color Online) a) Calculated magnetization for vanadium (black stars), chromium (blue squares), and manganese (red circles) substituted graphene. b) The calculated total magnetization for the magnetically substituted graphene. c) The magnetic moment for each atom between chromium atoms with the configuration for chromium substituted graphene separated by five carbon atoms.}
\label{mag3}
\end{figure}

Figure \ref{mag3}(a) illustrates the absolute magnetic moment on each magnetic atom in their respective configurations. As shown, vanadium, chromium, and manganese has an average moment of around 1, 3.5, and 3 $\mu_B$ respectively. Here, $\mu_B$ is the Bohr magneton. Figure \ref{mag3}(b) shows the total magnetic moment for the entire supercell structure. Here, the vanadium case stays constant at about 2 $\mu_B$, which is mainly produced by the vanadium atoms maintaining a constant ferromagnetism. However, the net magnetic moment of the entire lattice structure tends to fluctuate for chromium and manganese as the magnetic atoms flip between FM and AFM ground states. Figure \ref{mag3}(c) gives, for the example case of chromium, the magnetic moment on each of the connector carbons, where the inset details the pathway within the lattice.  Therefore, there is an induced magnetic moment in the carbon, which means the simple super-exchange between the magnetic atoms is due to a complex network of interactions throughout the carbon atoms.

\subsection*{Electronic structure and magnetization mapping}

\begin{figure*}
\includegraphics[width=1.0 \linewidth]{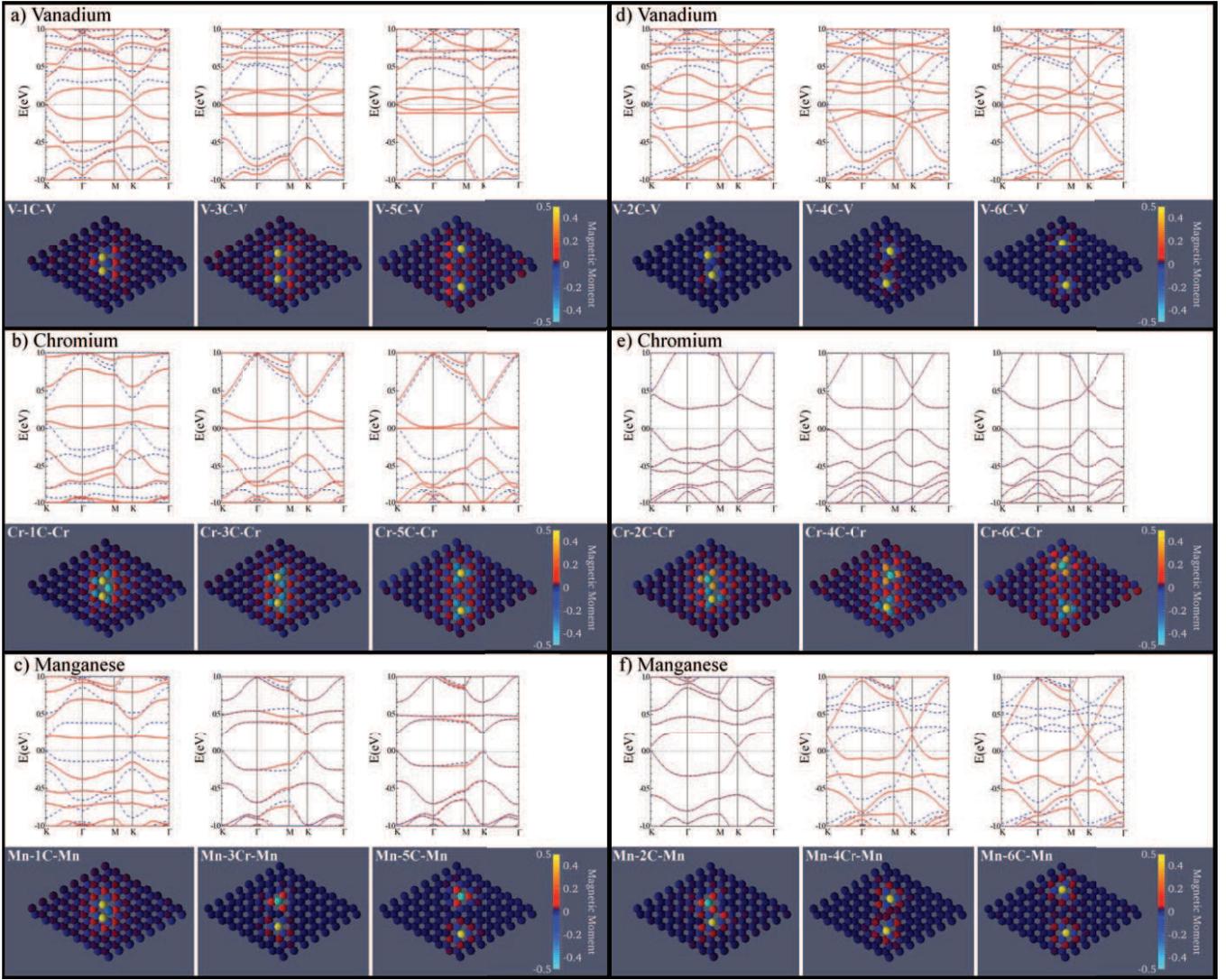}
\caption{(Color Online)  The electronic band structure and configuration mapped magnetic moment in magnetically-substituted graphene with odd (1(a), 3(b), and 5(c)) and even (2(d), 4(e), and 6(f)) separations between two vanadium (a,d), chromium (b,e), and manganese (c,f) atoms. The magnetic moments of the magnetic atoms are not given by the scale. They have magnetic moments of 1, 3, and 3.5 $\mu_B$ for vanadium, manganeses, and chromium, respectively.}
\label{mag}
\end{figure*}

Figure \ref{mag} shows the electronic band structure for each ground state configuration of graphene with vanadium (a,d), chromium (b,e), and manganese (c,f) substituted atoms. The lower panels detail the lattice configuration and the magnetic moment for each atom in the supercell, where the magnetic atoms are denoted as the extreme cases. Overall, these panels illustrate the increased position of the magnetic atom in relation to each other by either 1, 3, and 5 carbon atoms (Figure \ref{mag}(a-c)) or 2, 4, and 6 atoms (Figure \ref{mag}(d-f)). This figure illustrates the competition between the magnetic interactions between magnetic atoms and the carbon lattice. 

With respect to the electronic bandstructure, the competition between the interactions of the magnetic atoms and carbon atoms have distinct consequences on the presence of Dirac cone, which is the staple characteristic of pure graphene. It is clear that all of the odd carbon spacing (1, 3, and 5 carbon separations shown in Fig. \ref{mag}(a-c)) configurations seem to dissolve the Dirac cone. However, the Dirac cone seems to be restored within the spin down channels only when there exists an even carbon spacings (2, 4, and 6 carbon separations shown in Fig. \ref{mag}(d-f)) and a FM ground state between magnetic atoms. In the even cases for chromium (Figure \ref{mag}(e)), the AFM character of the chromium-chromium interaction enables the whole sheet to be AFM. Therefore, the spin up bands are degenerate with the spin down bands. However, this highlights the competition between various interactions in the superlattice.

The competition between M-C and C-C interactions seems to play a distinct role in the magnetic character for these configurations. In the vanadium cases, the magnetic atoms maintain an FM ground state throughout, which forces the C-C interactions in graphene lattice into either AFM and FM interactions. However, in the chromium cases, the C-C interactions between the magnetic atoms remain AFM, and the Cr-C interaction flips between FM and AFM. Therefore, the overall character of the Cr-Cr interaction changes and produces an RKKY-like fluctuation in the super-exchange. With manganese, the interactions are not trivial to determine.


\section*{Discussion}

As we examine magnetically-substituted graphene, we find that there are complex interactions between the electronic, magnetic, and lattice degrees of freedom within the 2D structure. Within the electronic structure, the notable feature of a Dirac cone in pure graphene (shown in Fig. \ref{gen-sub}(b)) seems to disappear in most cases. However, when the magnetic atoms are separated by an even number of carbon atoms and interacting with FM character (as shown in Fig. \ref{mag}(d) and (f)), the Dirac cone seems to be restored within the spin down channel. From our calculations of the magnetic moment for each atom, it is clear that the carbon atoms within proximity to the magnetic atoms gain a distinct magnetization, as opposed to the zero magnetic moment of the pure graphene lattice. Therefore, the presence of a magnetic moment disrupts the electronic structure through a competition between valence (those used for magnetism) and conduction (those used for electronic transfer) electrons. However, when the carbon atoms, between the two magnetic atoms, have a FM character, that frustration seems to be released and the graphene lattice is able to restore the Dirac cone. We can determine that the Dirac cone is produced by carbon electron interactions because the Dirac cone is only restored in the spin down electron spectrum, which does not include the magnetic atoms. The Dirac cone details a distinct complexity in the competition between magnetic and electronic order in the superlattice. Although the significance of this phenomena is of technical nature, the present calculations do present a clear visualization of a Friedel-like spin density variation of induced moments on carbon atoms. We note that we have also calculated the case of non-magnetic titanium atom substitution, no spin polarization is observed in the system. Instead, a Friedel oscillation in the charge density is obtained.

From Fig. \ref{exch}(b), we determine that the super-exchange between the two magnetic atoms is mitigated through both the spin up and spin down conduction electrons. This produces an increase in the PDOS at the Fermi level and explains why the carbon atoms are gaining magnetic moment when they are closer to the magnetic atoms (Fig. \ref{mag3}(c)). Here, the sharing of electrons between the magnetic atoms and the lattice produces a net magnetic moment on the carbon sites. Based on the calculated Mulliken populations, we can infer that spin states for the magnetic atoms are Cr$^{3+}$ ($S$ = 3/2), Mn$^{3+}$ (high-spin state $S$ = 1), and V$^{4+,5+}$ ($S$ = 1/2 or $S$ = 0) produced mainly from sharing three valance electrons with the graphene lattice. Vanadium is hard to gauge, since V$^{3+}$ should have an $S$ = 1 spin state. Therefore, we attribute its loss of magnetic moment to the possibility of a mixed valence state of V$^{4+}$ and V$^{5+}$ given a net magnetic moment of close to $S$ = 1/2 (shown in Fig. \ref{mag3}(a) and (b)).

Therefore, the addition of magnetic atoms produces a net magnetic moment in the graphene lattice around the atoms, which leads to better conduction and an overall super-exchange. This exchange can be quantified by examining the change in the total energy of the FM ($\uparrow \uparrow$) state and the AFM ($\uparrow \downarrow$) state. Using a standard Heisenberg Hamiltonian for an independent dimer
\be
{\cal H} = { J}\, \bm{ S}_{1}\cdot\bm{S}_{2},
\label{dimerH}
\ee
where $S_i$ is the quantum mechanical spin for sites $i$ = 1,2, it can be determined that the change in energy ($\Delta E$) from the FM and AFM state is simply proportional to the exchange energy $J$\cite{hara:05,hara:09}. Therefore, through an examination of the change in total energy of the two configurations, the super-exchange energy can be computationally estimated (shown in Fig. \ref{exch}(a)). This describes only the exchange between the two magnetic atoms. While there is an individual exchange between the induced magnetic moments on the carbon atoms, the interaction strength cannot be determined and can only be inferred as FM or AFM, based on the spin up or down nature of the magnetization. 

Depending on the coupling between the magnetic atoms and the carbons, magnetic frustration could be the cause of the variation between the FM and AFM characterization. As shown in Fig. \ref{mag}(a)-(d), vanadium maintains FM character as the magnetic atoms are separated, which the local magnetic moments on the carbons in between orient themselves in response. This indicates that $J_{V-C}$ (the exchange between vanadium and carbon) is dominant over $J_{C-C}$ (the exchange between carbon atoms). Therefore, vanadium is able to force the carbon moments into a configuration that works to maintain $J_{V-V} > 0$.

This effect is different for chromium (Fig. \ref{mag}(b)-(e)) and manganese (Fig. \ref{mag}(c)-(f)). In the chromium cases, $J_{C-C}$ seem to dominate $J_{Cr-C}$, since the $J_{Cr-Cr}$ alternates through between FM and AFM, which is similar for manganese. These interactions seems to indicate the presence of an RKKY exchange through the graphene lattice. However, in order to have an RKKY interaction, the graphene needs to have a metallic DOS at the Fermi level, which is established in Fig. \ref{exch}(b) by a distinct increase in the PDOS at the Fermi level for the carbon atoms between the magnetic atoms. Further work is being performed to try to characterize these interactions as well as the placement of other magnetic atoms. 

Overall, these calculations find a distinct RKKY exchange interaction. Since vanadium, chromium, and manganese are small transition-metal elements, the spin-orbit coupling is weak. The presence of spin-orbit coupling can produce a gap in the system forcing a standard ferromagnetic response. This was first investigated by Dugaev $et~al.$ using a standard exchange model with the spin-orbit coupling\cite{duga:06}, where they found that RKKY interactions should not be present due to this energy gap at the Fermi level. However, since we are using $3d$ transition-metal atoms, spin-orbit coupling can be assumed to be negligible. Therefore, we find that the presence of magnetic transition-metal atoms produces multiple electron bands at the Fermi level beyond the standard Dirac cone crossing, and that there are distinct correlations between magnetic atoms (as shown in Fig. \ref{mag}). 

\section*{Conclusion}

In this article, we present a comparative computational study examining the changes in the electronic and magnetic properties of magnetically-substituted graphene. Starting from a 128-atom supercell of graphene, two carbon atoms are then replaced with either vanadium, chromium, or manganese. The magnetic impurities are first considered with one carbon in between (1C) and separated until there are six carbons in between (6C) the magnetic atoms. Using a generalized-gradient approximation density functional theory with a full geometry optimization and spin polarization, we performed calculations to determine the total energy ground state, electronic band structure, density of states (DOS), and magnetic moment for each configuration. 

Through an analysis of the total energy for the different magnetic configurations, the super-exchange interaction between magnetic atoms is shown to alternate between antiferromagnetic and ferromagnetic state as a function of impurity distance. Additionally, we examine the spatially dependent magnetic moment for each atom in the superlattice and discover that the carbon atoms within close proximity to the magnetic atoms gain a distinct magnetic moment. The magnetic carbon atoms produce a complex network of exchange interactions. Through an analysis of the partial density of states for the carbon atoms, a distinct increase in electron mobility at the Fermi level is shown. Therefore, the calculation of a variable super-exchange with spatial distance and the transition to a more metallic state for the carbons in between the magnetic atoms indicates the presence of an RKKY-like super-exchange interaction through the graphene atoms.

Further investigation is needed to understand the induced magnetic moment in carbon and the exchange network between the magnetic atoms. Given the advancements in scanning tunneling microscopy (STM) and nanoscopic enhancements of atomic substitutions in graphene, experimental exploration of these interactions is a distinct possibility. The presence of variable magnetic exchange due to in-situ substitution and induced magnetism in graphene shows the increased possibility of using graphene for spintronic materials. These calculations are meant as the motivation for the use of magnetically-doped graphene and other 2D materials for the possible advancement for spintronic devices and quantum computation.

\section*{Computational Methods}

Using Atomistix Toolkit\cite{quantumwise,bran:02,sole:02}, we performed density functional theory (DFT) calculations within a spin-polarized generalized gradient approximation (SGGA) of PBE (Perdew, Burke, and Ernzerhof) functionals\cite{perd:96} for a 2D hexagonal lattice of graphene supercell (Fig. \ref{gen-sub}(a)) with two magnetic impurities substituted such that they were separated by N carbon atoms through minimum distance, where N = 1 - 6 (Fig. \ref{gen-sub}(c)). The lattice consisted of a total of 128 atoms and geometry was optimized using a force minimization method within an initial k-point sampling of 1x1x1. Further analysis was performed with a larger 3x3x1 k-point sampling once the geometry was minimized. The use of the smaller $k$-sampling was for purposes of computational time given that extent and number of calculations being performed. Geometry optimizations performed with the higher $k$-point sampling showed no significant difference.

We implemented initial spin states of up-up ($\uparrow \uparrow$), up-down ($\uparrow \downarrow$), and down-down ($\downarrow \downarrow$) configurations on the magnetically-substituted atoms and zero spin on the carbon atoms, which was established through non-substituted test runs on pure graphene. As expected, the configurations of $\uparrow \uparrow$ and $\downarrow \downarrow$ were found to be degenerate. In all, we perform calculations on 54 total configurations (6 spatial configurations, 3 magnetic configurations, 3 different magnetic atoms). The DFT calculations were run using a tolerance of 0.3 $\mu$eV with an energy cut off of 2 keV. As shown in Santos et al.\cite{sant:10}, the presence of a Hubbard U didn't show any major differences for the smaller magnetic atoms of vanadium, chromium, and manganese. This is discussed in the supplementary materials, but is most likely due to the individual presence of the atoms. Therefore, since we are doing a comparative study, we did not implement an Hubbard onsite potential. It should be noted that while the presence of a Hubbard U would restrict some electron movement by constraining the electronic orbitals, the general effects the trends for magnetization and electron conduction will still remain.

In all, we calculated the total energy of each system to find the ground state configuration assuming collinear polarized spins and estimated the exchange parameters between the two magnetic atoms. Furthermore, we determined the electronic band structure and density of states (DOS) to investigate the magnetic effect on the graphene Dirac cone at the $K$ point in the Brillouin zone, and we also map the induced magnetic moments for each atom to examine the interactions throughout the graphene supercell. 

\section*{Acknowledgements}

C.B.C., C.C., and J.T.H thank the support of the Tickle Summer Fellowship Program. The work at Los Alamos National Laboratory was carried out under the auspice of the U.S. DOE and NNSA under Contract No. DEAC52-06NA25396 and supported by U.S. DOE Basic Energy Sciences Office (T.A. and A.V.B.). Work of A.V.B. was also supported by European Research Council (ERC) DM 321031.This work was also, in part, supported by the Center for Integrated Nanotechnologies, a U.S. DOE Office of Basic Energy Sciences user facility (J.-X.Z).

\section*{Author Contributions}

J.T.H. and C.C. formulated the project. C.B.C ran all simulations and prepared all figures. J.T.H and C.B.C wrote the main text. C.C, T. A., J.-X. Z., and A.V. B. assisted on data analysis and the preparation of the manuscript. All authors reviewed the manuscript.

\section*{Additional Information}

The authors declare no competing financial interests. 


\begin{thebibliography}{harald}

\bibitem{wolf:01} Wolf, S. A. $et~al.$, Spintronics: A Spin-Based Electronics Vision for the Future. {\it Science} {\bf 294}, 1488-1495 (2001).

\bibitem{bade:10} Bader, S. D. \& Parkin, S. S. P. Spintronics. {\it Annual Review of Condensed Matter Physics} {\bf 1}, 71-88 (2010).


\bibitem{zuti:04}Zutic, I., Fabian, J. \& Das Sarma, S. Spintronics: Fundamentals and applications. {\it Rev. Mod. Phys.} {\bf 76}, 323-410 (2004).

\bibitem{roch:05}Rocha, A. R. et al. Towards molecular spintronics. {\it Nat. Mater.} {\bf 4}, 335-339 (2005).

\bibitem{boga:08} Bogani, L. \& Wernsdorfer, W. Molecular spintronics using single-molecule magnets. {\it Nat. Mater.} {\bf 7}, 179-186 (2008).


\bibitem{khaj:11} Khajetoorians, A. A., Wiebe, J., Chilian, B. \& Wiesendanger, R. Realizing All-SpinÐBased Logic Operations Atom by Atom. {\it Science} {\bf 332}, 1062-1064 (2011).

\bibitem{enge:01} Engel, H.-A., Recher, P. \& Loss, D. Electron spins in quantum dots for spintronics and quantum computation. {\it Solid State Communications} {\bf 119}, 229-236 (2001).

\bibitem{awsc:02} Awschalom, D., Loss, D. \& Samarth, N. Semiconductor Spintronics and Quantum Computation. (Springer Science \& Business Media, 2002).

\bibitem{awsc:07} Awschalom, D. D. \& Flatt\'e, M. E. Challenges for semiconductor spintronics. {\it Nat. Phys.} {\bf 3}, 153-159 (2007).


\bibitem{xu:06} Xu, Y. \& Thompson, S. Spintronic Materials and Technology. (CRC Press, 2006).

\bibitem{mats:03} Matsumoto, Y. et al. Combinatorial Investigation of Spintronic Materials. {\it MRS Bulletin} {\bf 28}, 734-739 (2003).

\bibitem{sato:02} Sato, K. \& Katayama-Yoshida, H. First principles materials design for semiconductor spintronics. {\it Semicond. Sci. Technol.} {\bf 17}, 367 (2002).

\bibitem{xiao:12} Xiao, D., Liu, G.-B., Feng, W., Xu, X. \& Yao, W. Coupled Spin and Valley Physics in Monolayers of MoS$_2$ and Other Group-VI Dichalcogenides. {\it Phys. Rev. Lett.} {\bf 108}, 196802 (2012).

\bibitem{saha:10} Saha, S. K., Baskey, M. \& Majumdar, D. Graphene Quantum Sheets: A New Material for Spintronic Applications. {\it Adv. Mater.} {\bf 22}, 5531-5536 (2010).

\bibitem{boehm:62} Boehm, H. P., Clauss, A., Fischer, G. O. \& Hofmann, U. Das Adsorptionsverhalten sehr dŸnner Kohlenstoff-Folien. {\it Z. anorg. allg. Chem.} {\bf 316}, 119-127 (1962).

\bibitem{geim:07} Geim, A. K. \& Novoselov, K. S. The rise of graphene. {\it Nat. Mater.} {\bf 6}, 183-191 (2007).

\bibitem{neto:09} Castro Neto, A. H. et al. The electronic properties of graphene. {\it Rev. Mod. Phys.} {\bf 81}, 109-162 (2009).


\bibitem{chen:08} Chen, J.-H., Jang, C., Xiao, S., Ishigami, M. \& Fuhrer, M. S. Intrinsic and extrinsic performance limits of graphene devices on SiO2. {\it Nat. Nano.} {\bf 3}, 206-209 (2008).

\bibitem{pop:12} Pop, E., Varshney, V. \& Roy, A. K. Thermal properties of graphene: Fundamentals and applications. {\it MRS Bulletin} {\bf 37}, 1273-1281 (2012).

\bibitem{bolo:08} Bolotin, K. I. et al. Ultrahigh electron mobility in suspended graphene. {\it Solid State Communications} {\bf 146}, 351-355 (2008).

\bibitem{moro:08} Morozov, S. V. et al. Giant Intrinsic Carrier Mobilities in Graphene and Its Bilayer. {\it Phys. Rev. Lett.} {\bf 100}, 016602 (2008).

\bibitem{bala:08} Balandin, A. A. et al. Superior Thermal Conductivity of Single-Layer Graphene. {\it Nano Lett.} {\bf 8}, 902-907 (2008).

\bibitem{zhu:10} Zhu, Y. et al. Graphene and Graphene Oxide: Synthesis, Properties, and Applications. {\it Adv. Mater.} {\bf 22}, 3906-3924 (2010).

\bibitem{lee:08} Lee, C., Wei, X., Kysar, J. W. \& Hone, J. Measurement of the Elastic Properties and Intrinsic Strength of Monolayer Graphene. {\it Science} {\bf 321}, 385Ð388 (2008).

\bibitem{sava:12} Savage, N. Materials science: Super carbon. {\it Nature} {\bf 483}, S30ÐS31 (2012).

\bibitem{ahme:14} Ahmed, T., Haraldsen, J. T., Zhu, J.-X. \& Balatsky, A. V. Next-Generation Epigenetic Detection Technique: Identifying Methylated Cytosine Using Graphene Nanopore. {\it J. Phys. Chem. Lett.} {\bf 5}, 2601Ð2607 (2014).

\bibitem{wehl:14} Wehling, T. O., Black-Schaffer, A. M. \& Balatsky, A. V. Dirac materials. {\it Advances in Physics} {\bf 63}, 1-76 (2014).

\bibitem{novo:05} Novoselov, K. S. et al. Two-dimensional gas of massless Dirac fermions in graphene. {\it Nature} {\bf 438}, 197-200 (2005).

\bibitem{yazy:08} Yazyev, O. V. \& Katsnelson, M. I. Magnetic Correlations at Graphene Edges: Basis for Novel Spintronics Devices. {\it Phys. Rev. Lett.} {\bf 100}, 047209 (2008).

\bibitem{rao:12} Rao, C. N. R., Matte, H. S. S. R., Subrahmanyam, K. S. \& Maitra, U. Unusual magnetic properties of graphene and related materials. {\it Chemical Science} {\bf 3}, 45 (2012).


\bibitem{pesi:12} Pesin, D. \& MacDonald, A. H. Spintronics and pseudospintronics in graphene and topological insulators. {\it Nat. Mater.} {\bf 11}, 409Ð416 (2012).

\bibitem{sant:10} Santos, E. J. G., Ayuela, A. \& S‡nchez-Portal, D. First-principles study of substitutional metal impurities in graphene: structural, electronic and magnetic properties. {\it New J. Phys.} {\bf 12}, 053012 (2010).

\bibitem{sher:11} Sherafati, M. \& Satpathy, S. RKKY interaction in graphene from the lattice GreenÕs function. {\it Phys. Rev. B} {\bf 83}, 165425 (2011).

\bibitem{hu:11} Hu, F. M., Ma, T., Lin, H.-Q. \& Gubernatis, J. E. Magnetic impurities in graphene. {\it Phys. Rev. B} {\bf 84}, 075414 (2011).

\bibitem{wang:15} Wang, Z., Tang, C., Sachs, R., Barlas, Y. \& Shi, J. Proximity-Induced Ferromagnetism in Graphene Revealed by the Anomalous Hall Effect. {\it Phys. Rev. Lett.} {\bf 114}, 016603 (2015).

\bibitem{gome:12} Gomes, K.K., Mar, W., Ko, W., Guinea, F., \& Manoharan, H.C. Designer Dirac fermions and topological phases in molecular graphene. {\it Nature} {\bf 483}, 306 (2012).

\bibitem{wang:12} Wang, H. et al. Doping Monolayer Graphene with Single Atom Substitutions. {\it NanoLetters} {\bf 12}, 141-144 (2012).

\bibitem{lv:12} Lv, R. et al. Nitrogen-doped graphene: beyond single substitution and enhanced molecular sensing. {\it Sci. Reports} {\bf 2}, 586 (2012).

\bibitem{duga:06} Dugaev, V. K., Litvinov, V. I. \& Barnas, J. Exchange interaction of magnetic impurities in graphene. {\it Phys. Rev. B} {\bf 74}, 224438 (2006).


\bibitem{hara:05} Haraldsen, J. T., Barnes, T. \& Musfeldt, J. L. Neutron scattering and magnetic observables for S=1/2 spin clusters and molecular magnets. {\it Phys. Rev. B} {\bf 71}, 064403 (2005).


\bibitem{hara:09} Haraldsen, J. T. \& Fishman, R. S. Spin rotation technique for non-collinear magnetic systems: application to the generalized Villain model. {\it J. Phys.: Condens. Matter} {\bf 21}, 216001 (2009).

\bibitem{rude:54} Ruderman, M. A. \& Kittel, C. Indirect Exchange Coupling of Nuclear Magnetic Moments by Conduction Electrons. {\t Phys. Rev.} {\bf 96}, 99-102 (1954).


\bibitem{kasu:56} Kasuya, T. A, Theory of Metallic Ferro- and Antiferromagnetism on ZenerÕs Model. {\it Prog. Theor. Phys.} {\bf 16}, 45-57 (1956).

\bibitem{yosi:57} Yosida, K., Magnetic Properties of Cu-Mn Alloys. {\it Phys. Rev.} {\bf 106}, 893-898 (1957).

\bibitem{neto:09} Castro Neto, A. H. \& Guinea, F. Impurity-Induced Spin-OrbitCoupling in Graphene. {\it Phys. Rev. Lett.} {\bf 103}, 026804 (2009).

\bibitem{sing:09} Singh, R. \& Kroll, P. Magnetism in graphene due to single-atom defects: dependence on the concentration and packing geometry of defects. {\it J. Phys.: Condens. Matter} {\bf 21}, 196002 (2009).

\bibitem{quantumwise} Atomistix ToolKit version 13.8, QuantumWise A/S (www.quantumwise.com)

\bibitem{bran:02} Brandbyge, M., Mozos, J.-L., Ordej\'on, P., Taylor, J. \& Stokbro, K. Density-functional method for nonequilibrium electron transport. {\it Phys. Rev. B} {\bf 65}, 165401 (2002).

\bibitem{sole:02} Soler, J. M. et al. The SIESTA method for ab initio order-N materials simulation. {\it J. Phys.: Condens. Matter} {\bf 14}, 2745 (2002).

\bibitem{perd:96} Perdew, J. P., Burke, K. \& Ernzerhof, M. Generalized Gradient Approximation Made Simple. {\it Phys. Rev. Lett.} {\bf 77}, 3865-3868 (1996).



\end{thebibliography}
\end{document}